\documentstyle[12pt,epsf]{article}
\newcommand{\br}{{\bf r}}
\newcommand{\half} {\frac{1}{2}}
\newcommand{\nk} {n_k}
\newcommand{\lk} {l_k}
\newcommand{\jk} {j_k}
\newcommand{\mk} {m_k}
\newcommand{\tk} {t_k}
\newcommand{\nsi} {n_i}
\newcommand{\li} {l_i}
\newcommand{\ji} {j_i}
\newcommand{\mi} {m_i}
\newcommand{\ti} {t_i}
\newcommand{\bsigma}{\mbox{\boldmath $\sigma$}}
\newcommand{\btau}{\mbox{\boldmath $\tau$}}
\newcommand{\threej}[6]{ \left( \begin{array}{ccc}
                               #1 & #2 & #3 \\
                               #4 & #5 & #6 
                             \end{array}
                        \right) } 
\newcommand{\sixj}[6]{ \left\{ \begin{array}{ccc}
                               #1 & #2 & #3 \\
                                #4 & #5 & #6 
                               \end{array}
                        \right\} } 
\newcommand{\ninej}[9]{ \left\{ \begin{array}{ccc}
                                 #1 & #2 & #3 \\
                                 #4 & #5 & #6 \\
                                 #7 & #8 & #9 
                                \end{array}
                         \right\} } 

\begin{document}
\centerline{\large {\bf Tensor interaction in Hartree--Fock calculations}}

\vspace{1.5cm}

\centerline{\bf Giampaolo Co'}
\centerline{Dipartimento di Fisica, Universit\`a di Lecce}
\centerline{and I.N.F.N. sezione di Lecce, I-73100 Lecce, Italy}

\vspace{.5cm}

\centerline{\bf Antonio M. Lallena}
\centerline{Departamento de F\'{\i}sica Moderna, Universidad de Granada,}
\centerline{E-18071 Granada, Spain}

%\noindent

%%%%%
%%%%%\maketitle
%
\vspace{1.5cm}

\begin{abstract}
The contribution of the tensor interaction in Hartree--Fock
calculation for closed shell nuclei is studied. 
The investigation is done neglecting the spin--orbit component of the
force. As expected from nuclear matter estimates
we find that the effect of the tensor interaction
is negligible in the case of spin--isospin saturated
nuclei and also for medium and heavy nuclei with closed
shells. In light nuclei with only one of the two spin--orbit 
partners filled, such as $^{12}$C and $^{14}$C, the tensor interaction
plays an important role.

\end{abstract}
\vskip 0.5 cm
PACS {21.60.Jz}
%
%x
\newpage
\section{Introduction}
One of the peculiarities of the nucleon--nucleon interaction is
the presence of non central terms. Already at the beginning of
the 40's the existence of a quadrupole moment of the deuteron
was recognized \cite{kel39}. This fact was explained by Rarita 
and Schwinger in terms of a static tensor force \cite{rar41}.
The presence of tensor terms is an essential ingredient in the
modern phase--shift analysis of nucleon--nucleon scattering data
\cite{ber90}. In a meson exchange picture of the
nucleon--nucleon interaction, tensor terms arise already when
the exchange of the lightest and better known meson, the pion,
is considered.

The relevance of the tensor part of the nucleon-nucleon
interaction is not only restricted to two--body systems.
Calculations based on microscopic nucleon--nucleon interactions of
light and  medium--heavy nuclei \cite{fab97} 
and nuclear matter \cite{wir88} show that binding is obtained only
because of the presence of the tensor force.

Although the tensor terms of the interactions are essential in
microscopic calculations, they are usually neglected in
effective theories where the Schr\"odinger equation is solved at
the mean--field level, since their contribution is though to be small.
This attitude is not unreasonable if one
consider that the tensor terms of the interactions do not contribute
in mean--field calculations of the binding energy of symmetric nuclear
matter.  On the other hand, 
since the spin and isospin structure of finite
nuclei is richer than that of nuclear matter, one may expect to
find, in the former systems some observable sensitive to the
presence of the tensor terms of the interaction.  In this
respect Random Phase Approximation calculations have shown that
magnetic excitations can be reasonably described only when
tensor terms are included in the effective interaction
\cite{spe80}.

We have investigated the effects of the tensor terms of the
nucleon-nucleon interaction on ground state properties of
doubly--magic nuclei and in this work we would like to answer 
the question whether it is reasonable to neglect this part of
the interaction in Hartree--Fock calculations. 
We started our study by using
a well know finite range interaction, the Brink and
Boeker B1 force \cite{bri67}, defined only in the four central
components. To this basic interaction we have added the tensor and
tensor--isospin terms taken from the realistic Argonne $v_{14}$
potential \cite{wir84} and from the phenomenological J\"ulich--Stony
Brook (JSB) interaction \cite{spe80}. Our investigation consists in
comparing the results obtained with these three different
interactions for various closed shell nuclei.

We briefly describe the formalism in the next section and we present 
the results of our investigation in Sect.  
\ref{results}. We shall show and explain the fact 
that the contribution of the tensor
interaction is negligible for those nuclei where all the spin--orbit
partners single particle levels are occupied. For the other nuclei, these
contributions remains small if the nuclei are heavy.

\section{The formalism}
The Hartree--Fock (HF) equations are obtained 
searching for a minimum of the energy functional
within a Hilbert space restricted to states which are Slater
determinant of single particle (sp) wave functions. In this way the
many--body problem is transformed in a set of many one--body problems.
For hamiltonians containing only two--body interactions, without
explicit density dependent terms, one has to solve, in coordinate space,
a set of equations of the type:
\begin{equation}
\label{hf1}
-\frac {\hbar^2}{2m_k} {\nabla_1}^2 \phi_k({\bf r}_1)\,
+ \, U({\bf r}_1)\, \phi_k({\bf r}_1) \,
- \, \int \, {\rm d}^3r_2 \, W({\bf r}_1,{\bf r}_2) \,
\phi_k({\bf r}_2) \, = \, \epsilon_k \phi_k({\bf r}_1) \, ,
\end{equation}
where we have indicated with $k$ the set of quantum
numbers characterizing the sp wave function
$\phi_k$ and  the sp energy $\epsilon_k$. The two
quantities $U$ and $W$ are the standard Hartree and Fock--Dirac
potentials \cite{rin80}

The only input of the theory is the effective nucleon--nucleon
interaction which we have chosen as a six components finite--range
two--body force of the type: 
\begin{equation}
\label{force1}
V(\br_1,\br_2) \, = \, \sum_{p=1}^6 \, V_p(\br_1,\br_2) \, O_p(1,2)
\end{equation}
where $O_p(1,2)$ indicates the following operators $1$,
$\btau_1 \cdot \btau_2$,
$\bsigma_1 \cdot \bsigma_2$,
$\bsigma_1 \cdot \bsigma_2 \,\, \btau_1 \cdot \btau_2$, $S_{12}$,
$ S_{12} \btau_1 \cdot \btau_2$,
for $p=1, \ldots ,6$, respectively. In this
expression we have adopted the usual convention for the  tensor operator:
\begin{equation}
S_{12}\, = \, 3\, 
\frac{ (\bsigma_1 \cdot \br_{12})(\bsigma_2 \cdot \br_{12})}
   { (\br_{12})^2} \, - \, \bsigma_1 \cdot \bsigma_2 \, ,
\end{equation}
with $\br_{12}=\br_1-\br_2$.

We suppose that all the functions $V_p$ in eq. (\ref{force1}) depend 
only from the relative distance between the two particles, 
$|\br_{12}|$. It is therefore convenient to use expressions of the 
sp wave functions  where the angular and
radial coordinates are separated:
\begin{eqnarray}
\label{spwf}
\phi_k(\br) & \equiv & \phi^t_{nljm}(\br)\, \equiv \, 
\frac{u^t_{nlj}(r)}{r} \, | ljmt \rangle \nonumber \\
& \equiv & \frac{u^t_{nlj}(r)}{r} \, 
\sum_{\mu s} \, \langle l \mu \half s | j m \rangle \,
 Y_{l\mu}(\theta,\varphi) \, \chi_s \, \chi_t \, .
\end{eqnarray} 
In the above expression $\theta$ and $\varphi$ are
the angular coordinates, $Y_{l\mu}$ is a spherical harmonics, 
$\chi_s$ and $\chi_t$ are the spin and isospin wave functions,
$s$ and $t$ represent the third components of spin and isospin,
and $\langle l \mu \half s | j m\rangle$ is a 
Clebsh--Gordan coefficient.

Multiplying eq. (\ref{hf1}) by  
$\langle \lk \jk \mk \tk|$ and
integrating on the angular, spin and isospin variables we obtain:
\begin{eqnarray}
\label{hf2}
-\frac{\hbar^2}{2m_k}
\left(\frac{d^2}{dr^2}-\frac{\lk(\lk+1)}{r^2}\right)
u^{\tk}_{\nk\lk\jk}(r) \,+ \,
U^{\tk}_{\nk\lk\jk}(r)u^{\tk}_{\nk\lk\jk}(r) \,
 - \, E^{\tk}_{\nk\lk\jk}(r)
\nonumber
\\
= \, \epsilon^{\tk}_{\nk\lk\jk}u^{\tk}_{\nk\lk\jk}(r) \, .
\end{eqnarray}

To evaluate $ U^{\tk}_{\nk\lk\jk}(r)$ and  $ E^{\tk}_{\nk\lk\jk}(r)$
we write the interaction
$V(\br_1,\br_2)$ in terms of its Fourier transform. This allows us 
to separate the functional dependence of the two coordinates $\br_1$ and
$\br_2$. 

Spin and isospin conservation implies that in the direct term only the
scalar (p=1) and the isospin (p=2) channels of the interaction
contribute. For these direct terms we find the expression:
\begin{equation}
\label{udirect}
\left[ U^{\tk}_{\nk\lk\jk}(r) \right] _{(p)} \, = \,
4 \sqrt{2\pi} \, \int \, {\rm d} r' \, r'^2 \, {\cal V}^p_0(r,r') 
\, \left[ {\cal U}_k(r') \right]_{(p)}
\, ,  
\end{equation}
where we have defined
\begin{equation}
{\cal V}^p_L(r,r') \, = \,
\int \, {\rm d}q \, q^2 \, j_L(qr)\, j_{L}(qr') \, V^{(0)}_p(q) \, ,
\end{equation}
with $j_L(x)$ a spherical Bessel function and
\begin{equation}
\displaystyle
V^{(0)}_p(q)\, = \, \sqrt{\frac{2}{\pi}} \, \int \, {\rm d}r \, r^2 \,
j_0(qr) \, V_p(r) \, .
\end{equation}
Finally, the function ${\cal U}$ is:
\begin{equation}
\label{uuu}
\left[ {\cal U}_k (r) \right]_{(p)} \, = \,
\left\{ \begin{array}{ll}
        \rho(r) \, , & p=1 \\
        2 \, \tk \, \left[ \rho^p(r)-\rho^n(r)\right] \, , & p=2
\end{array}
\right.
\end{equation}
with $\tk=\half$ ($-\half$) according to $k$ being a proton (neutron).

In eq. (\ref{uuu}) we have indicated with $\rho$ the nucleon density:
\begin{equation}
\label{density}
\rho(r) \, = \, 
\sum_{nljt} \, \frac{2j+1}{4\pi} \, 
\left( \frac{u^t_{nlj}(r)}{r} \right)^2 \, ,
\end{equation} 
and with $\rho^p$ and $\rho^n$ the proton and neutron densities,
defined as in eq. (\ref{density}) without the sum on $t$, such as
$\rho=\rho^p+\rho^n$. 

All the channels of the force
contribute to the exchange term in eq. (\ref{hf2}), which are expressed as: 
\begin{eqnarray}
\label{eee}
\left[ E^{\tk}_{\nk\lk\jk}(r)\right]_{(p)} \, = \,
\displaystyle
\sqrt{ \frac{2}{\pi} } \, \sum_{\nsi\li\ji\mi\ti} \, 
u^{\ti}_{\nsi\li\ji}(r) \, \int \, {\rm d}r' \, 
u^{\ti}_{\nsi\li\ji}(r') \, u^{\tk}_{\nk\lk\jk}(r') \nonumber \\
 \sum_L \, \xi(\lk+\li+L) \, (2 \ji+1) \, (2L+1) \, I_p \, \left[{\cal
E}_{kiL}(r,r')\right]_{(p)} \, ,
\end{eqnarray} 
where we have introduced the function 
\begin{equation}
\xi(l)=\half \left[ 1+ (-1)^l \right] \, .
\end{equation}
The term $I_p$ is obtained from the calculation of the isospin 
matrix elements and is given by:
\begin{equation}
I_p \, = \, \left\{ 
\begin{array}{ll}
\delta_{\tk, \ti} & p=1,3,5 \\
2\delta_{\tk,-\ti} +\delta_{\tk,\ti} & p=2,4,6
\end{array}
\right. \, ,
\end{equation}
where we have indicated 
with $\delta$ the Kronecker symbol. Finally, the function ${\cal E}$
includes the angular momentum coupling pieces. 
For the scalar and isospin terms we have the following expression:
\begin{eqnarray}
\label{escalar}
\left[ {\cal E}_{kiL}(r,r')\right]_{(p=1,2)} \, = \,
\displaystyle
\threej {\jk} {\ji} {L} {\half} {-\half} {0} ^2 \, {\cal V}^p_L(r,r') \,
,
\end{eqnarray} 
where we have made use of the Racah 3j symbol.

The contribution of the spin dependent terms is given by:
\begin{eqnarray}
\label{espin}
\left[ {\cal E}_{kiL}(r,r') \right]_{(p=3,4)} \, = \,
\displaystyle
2 \, \left[
\threej {\lk} {\li} {L} {0} {0} {0} ^2
-\half \threej {\jk} {\ji} {L} {\half} {-\half} {0} ^2 
\right] {\cal V}^p_L(r,r') \, .
\end{eqnarray} 

Finally, for the tensor channels we have:
\begin{eqnarray}
\label{etensor}
\left[ {\cal E}_{kiL}(r,r') \right]_{(p=5,6)} \, = 
\displaystyle
 2 \sqrt{30} \, (2 \li+1) \, (2 \lk+1) \, \sum_{L' J} \, 
(-i)^{L-L'} \, (-1)^J \, {\cal V}^p_{LL'}(r,r') \nonumber \\ 
 (2J+1) \, (2L'+1) \, 
\sixj {L} {L'} {2} {1} {1} {J}
\threej {L'} {L} {2} {0} {0} {0} \\  
\ninej {\lk} {\half} {\jk} {\li} {\half} {\ji} {L} {1} {J}
\threej {\lk} {\li} {L} {0} {0} {0}  
\ninej {\lk} {\half} {\jk} {\li} {\half} {\ji} {L'} {1} {J}
\threej {\lk} {\li} {L'} {0} {0} {0} \nonumber \, ,
\end{eqnarray} 
where the Racah 6j and 9j symbols have been used and where we have
introduced the function
\begin{equation}
{\cal V}^p_{LL'}(r,r') \, = \,
\int \, {\rm d}q \, q^2 \, j_L(qr)\, j_{L'}(qr') \, V^{(2)}_p(q) \, ,
\end{equation}
with 
\begin{equation}
\displaystyle
V^{(2)}_p(q)\, = \, -\sqrt{\frac{2}{\pi}} \, \int \, {\rm d}r \, r^2 \,
j_2(qr) \, V_p(r) \, .
\end{equation}

The solution of eq. (\ref{hf2}) provide us with the sp energies
$\epsilon$ and wave functions $u(r)$. This solution has been 
obtained iteratively using the plane waves expansion method of 
refs. \cite{gua82}.
The center of mass motion has been considered in its  simplest
approximation, consisting in inserting the nucleon reduced mass in the
hamiltonian. 
The sp wave functions used to start the iterative procedure have
been generated by a Saxon--Woods potential without spin-orbit and
Coulomb terms. This means we have used the same starting wave 
functions for both protons and neutrons
and for sp states which are spin--orbit partners.

The total energy of the system is calculated as:
\begin{eqnarray}
E \, = \,
\displaystyle
\sum_k \, \epsilon_k \, - \, \half \, \sum_{ki} \,
\left[ \int \, {\rm d}^3r_1 \, {\rm d}^3r_2 \,
\phi_k^*(\br_1) \, \phi_j^*(\br_2) \, V(|\br_{12}|) \,
\phi_k(\br_1) \, \phi_j(\br_2) \right. \nonumber
\\
  \left. \hspace{2cm} \displaystyle 
- \, \int \, {\rm d}^3r_1 \, {\rm d}^3r_2 \,
\phi_k^*(\br_1) \, \phi_j^*(\br_2) \, V(|\br_{12}|) \,
\phi_j(\br_1) \, \phi_k(\br_2) \right] 
\\
 = \displaystyle 
 \sum_{nljt} \, (2j+1) \, \epsilon^{t}_{nlj} \,
- \, \half \left[ 4\pi \int \, {\rm d}r \, r^2 \, \rho(r) 
\left(\sum_{p=1}^6 \left[ U^{t}_{nlj}(r)\right]_p \right) 
\right. \nonumber \\ 
\left . - \, \sum_{nljt} \, (2j+1)\, \int \, {\rm d}r 
\left(\sum_{p=1}^6 \left[ E^{t}_{nlj}(r)\right]_p \right)
\right]
\end{eqnarray}

\section{Results and discussion}
\label{results} 
Our calculations are based on a finite 
range effective force widely used in HF calculations, 
the B1 interaction of Brink and Boeker 
\cite{bri67} which was constructed to reproduce the experimental binding
energies of $^4$He, $^{16}$O and $^{40}$Ca. 
With  this interaction we have performed HF calculations also for 
$^{12}$C, $^{14}$C, $^{48}$Ca and $^{208}$Pb nuclei. 

A comparison between the binding energies obtained in our calculations 
with the experimental values \cite{aud93} 
is given in table \ref{tableb1}.
While the energies of  $^4$He, $^{16}$O and
$^{40}$Ca are rather well reproduced, the model fails 
in describing those of the other nuclei. 
A study of the contribution of the four components of the force shows
that only the spin--isospin (p=4) term provides attraction. 
The contribution of the isospin (p=2) and spin (p=3) channels is the
same for the $Z=N$ nuclei. Since in the present
calculations protons and neutrons sp wave functions are identical,
for $Z=N$ nuclei the proton and neutron densities are equal,
therefore the direct isospin term, eq. (\ref{uuu}), is zero and 
the two exchange terms for p=2 and p=3, give the same result. 
Of course this picture breaks down for nuclei with
neutron excess like $^{14}$C, $^{48}$Ca and $^{208}$Pb. 

To investigate the influence of the tensor component of the force on
HF calculations we have added to the B1 interaction the tensor
force obtained from the Argonne $v_{14}$ potential \cite{wir84} and
that obtained from the J\"ulich--Stony Brook (JSB) interaction
\cite{spe80}.
The tensor terms of the JSB interaction are
constructed by considering the exchange of the
$\pi$ plus $\rho$ mesons, therefore they are active only
in the tensor isospin channel. On the other hand,
the pure tensor term of the Argonne potential is very
small if compared with the tensor-isospin one. For this reason in our
calculations we have used only the tensor-isospin terms of 
the two interactions.
These terms are shown in the panel (a) of fig. \ref{fig1}.

This treatment of the tensor part of the interaction reproduces rather
well the tensor terms of an effective interaction calculated within a
Brueckner G-matrix approach. In ref. \cite{nak84} is it shown that the
effective force in the tensor channels is very similar to the bare
nucleon--nucleon interaction.
From fig. \ref{fig1} the similarity between the JSB and Argonne tensor isospin
channels is evident in the low momentum region.

We did not use the full Argonne and JSB interactions because
they are not suited for HF calculations. The Argonne potential is a
realistic nucleon--nucleon interaction which reproduces 
the experimental data of the two nucleon systems, and therefore it has
a strong repulsive core. With this interaction we have 
calculated $^{16}$O and $^{40}$Ca and both of them 
are unbound ($52.08$ and $217.18$ MeV respectively).
The JSB interaction is an effective force designed to describe
low--lying excited states within the Landau--Migdal theory of finite
Fermi systems, and for HF calculation it turns out to be too
attractive.  

The binding energies obtained with the
interactions constructed by adding to the B1 interaction the 
tensor terms of the Argonne $v_{14}$ potential (A) and that of the 
JSB interaction (JSB) are shown in table \ref{tableten}. 
A comparison with the results of table
\ref{tableb1} shows that the contribution of the tensor force is
very small.

The set of sp energies for the nuclei we have studied is shown in
tables \ref{tablec12}--\ref{tableca48}. For brevity we
do not show these energies for the $^{208}$Pb case.
One can notice that in the $^{16}$O and  $^{40}$Ca  nuclei, 
the sp energies of the spin--orbit partner levels are the same, while
for $^{14}$C and $^{48}$Ca they are different.

This results can be understood analyzing eq. (\ref{etensor}). If the
$r$--dependent terms would be constant, the angular momentum sums for 
nuclei having spin--orbit partner states occupied would be exactly
zero. 
In reality each term is identified by small variations produced by
the integral on $q$, depending from $L$ and $L'$ and the integral on $r$
depending from the sp wave functions. These variations 
are not large enough to prevent big cancelations
between the contribution of the spin--orbit partners terms. 
The contribution of the tensor interaction can be significant in nuclei
where not all the sp spin--orbit partners levels are occupied.
In closed shell nuclei there are at most two of such a levels, 
one for neutrons and one for protons. 
The contribution of the tensor
channel should be compared with that of the other channels where all
the sp states contribute. In light nuclei, like $^{12}$C, 
the tensor interaction could produce a noticeable contribution if compared
with that generated by the other channels of the force which involves
few sp levels. We expect that in heavier nuclei the tensor contribution
produced by only one sp level should be compared with that produced
by the other, many, sp levels for the central channels of the force.
For this reason in the tables \ref{tablec12}--\ref{tableca48} the
effect of the tensor force is noticeable only for those nuclei having
spin--orbit partners not fully occupied. 

It is interesting to notice that the spin--orbit splitting 
produced by the tensor force goes in the opposite direction with
respect to the experimental one. 
Clearly, a more realistic description of the
nuclear ground state in terms of mean field model requires the
presence of explicit spin--orbit terms in the interaction. 

In order to test the sensitivity of the $^{12}$C ground state properties
to the tensor interaction we have inserted a tensor--isospin channel
of gaussian shape (in coordinate space) whose parameters have been
fixed to reproduce the experimental binding energy of this nucleus. 
This new tensor force is shown in the  panel (b) 
of fig. \ref{fig1} and the results obtained with this new interaction for the
other nuclei are shown in the  tables  by  the 
columns labeled T6. It is worth to notice that this interaction has
opposite sign with respect to the tensor terms of the JSB and Argonne
interaction and a much stronger strength.

The comparison done in table \ref{tableten}
between the binding energies obtained with this new
interaction with those previously obtained shows that only the results
of the two carbon isotopes have been strongly modified.
In this table we see that the binding energies of the nuclei
with all spin--orbit partners levels occupied are very little modified
by the tensor force. The $^{48}$Ca and $^{208}$Pb nuclei show
more sensitivity to the modification of the tensor component of the
force, but it is however rather small. 
The same kind of effects can be seen in the other tables showing the
sp energies. The sp energies of  $^{12}$C and $^{14}$C nuclei are
strongly modified by the new tensor force, but those of
$^{16}$O and $^{40}$Ca remain the same. 
In $^{48}$Ca the modification of the sp energies are relevant, even
though the change on the binding energy is very small
($\sim$ 2\%).

In order to see how the modification of the tensor interaction can
affect the sp wave functions we have calculated the nucleon
distributions. As example of the results obtained
we show in fig. \ref{fig2}  the proton distributions of the
$^{12}$C, $^{14}$C, $^{40}$Ca and $^{48}$Ca nuclei. 
On the scale of the figure the curves representing the Argonne and JSB
results are overlapped to the curves of the B1 results.
The figure shows
that the JSB and Argonne tensor potentials do not produce sizeable
effects on the proton distributions. The situation is quite different
when the T6 tensor terms is used. The proton distribution of $^{12}$C
and  $^{14}$C are quite different, while the effect on  
$^{40}$Ca and $^{48}$Ca are negligible. These results confirm the
previous analysis on the behaviour of the tensor interaction. The
tensor force plays an important role in $^{12}$C and $^{14}$C, 
but its effects are 
essentially zero for nuclei with fully occupied spin--orbit partner
levels, and very small for medium--heavy nuclei.

\section{Conclusions}
\label{concl}
In this work we have studied the effects of the tensor interaction on HF
calculations. We found that there are big cancelations between
the contributions of sp levels which are spin--orbit partners. 
For this reason, only the contribution of those sp
states not having occupied the spin orbit partner level
is significant.
A closed shell nucleus can have at most two of these
states, one for protons and one for neutrons, and, as a consequence, 
the effect is relatively small 
in heavy nuclei where it has to be compared
with the effect of the other terms of the force active on all the sp 
levels. The situation is different for $^{12}$C where only two levels 
are occupied, therefore, as we have shown, the effect of the tensor
interaction is comparable with that of the other terms.

One may argue whether the difficulties in reproducing the $^{12}$C
properties of traditional HF calculations can be overcome by
inserting the tensor terms of the interaction.
We have shown that, if these terms are included, 
it is possible to reproduce the $^{12}$C binding energies
without spoiling the agreement obtained for the other nuclei. 
To obtain this results the use of tensor forces very different from
those of Argonne and JSB interactions has been necessary.
On the other hand, the study of excited  low-lying magnetic
states \cite{co90} gives strong indications
that the effective tensor isospin interaction should be quite similar
to those of the Argonne and JSB interactions. The unified description
of ground and excited nuclear states within a Hartree-Fock plus Random
Phase Approximation theoretical scheme, rules out the possibility of
fixing the tensor part of the effective interaction with a fit to the 
$^{12}$C ground state properties.
 
The task of constructing a new finite--range interaction to be 
used in mean field calculations was beyond the aim of the
present paper, and it will be pursued in the future, by adding finite
range spin--orbit and density dependent terms. In any case 
the results we have presented show the need of including  tensor
terms in this effective interaction if one wants the model to be valid
on the full isotope table.

%
%
%\newpage

\newpage
\begin{table}
\caption{Binding energies, in MeV, obtained with the B1 interaction.
The experimental values are taken from ref.
\protect\cite{aud93}.}
\label{tableb1}
\begin{center}
\begin{tabular}{crr}
\hline
           & exp     &   B1     \\ 
\hline
$^4$He     &  -28.30 &   -28.22 \\
$^{12}$C   &  -92.16 &   -48.68 \\
$^{14}$C   & -105.28 &   -74.49 \\
$^{16}$O   & -127.62 &  -113.55 \\
$^{40}$Ca  & -342.05 &  -340.75 \\
$^{48}$Ca  & -416.00 &  -362.82 \\
$^{208}$Pb &-1636.45 & -2059.55 \\
\hline
\end{tabular}
\end{center}
\end{table}
\begin{table}
\caption{Binding energies, in MeV, obtained with the interaction
obtained by adding to the B1 interaction
the tensor terms of ref. \protect\cite{spe80} (JSB) and that
of ref. \protect\cite{wir84} (A). The column labeled T6 shows the
results obtained with the tensor force adjusted to reproduce the
experimental value of the $^{12}$C binding energy.}
\label{tableten}
\begin{center}
%\begin{tabular}{|c|rrr|}
\begin{tabular}{crrr}
\hline
           & JSB     &   A      & T6\\ 
\hline
$^4$He     &  -28.22 &   -28.22 & -28.22  \\
$^{12}$C   &  -47.17 &   -47.46 & -92.36  \\
$^{14}$C   &  -74.56 &   -74.59 & -90.45  \\
$^{16}$O   & -114.10 &  -114.10 & -113.53 \\
$^{40}$Ca  & -342.40 &  -342.40 & -340.75\\
$^{48}$Ca  & -364.01 &  -363.98 & -377.01\\
$^{208}$Pb &-2074.51 & -2074.00 & -2063.93\\
\hline
\end{tabular}
\end{center}
\end{table}
\begin{table}
\caption{Single particle energies, in MeV, for $^{12}$C for the 
interactions used. Since these calculations have been done without the 
Coulomb interaction, proton and neutron single particle energies are 
the same.}
\label{tablec12}
\begin{center}
\begin{tabular}{crrrr}
\hline
 $^{12}$C       & B1      & JSB      &   A  & T6   \\ 
\hline
 1s$_{1/2}$     & -36.81  & -36.66 & -36.55 & -42.02 \\
 1p$_{3/2}$     & -11.82  & -11.29 & -11.36 & -25.82 \\
\hline
\end{tabular}
\end{center}
\end{table}
\begin{table}
\caption{Single particle energies, in MeV, for $^{14}$C. 
We have indicated with $\pi$ the proton and with $\nu$ the neutron
levels.}
\label{tablec14}
\begin{center}
\begin{tabular}{crrrr}
\hline
 $^{14}$C         & B1      & JSB      &   A  & T6    \\ 
\hline
 1s$_{1/2}$ $\pi$ & -44.83  & -44.93 & -44.93 & -44.24 \\
 1p$_{3/2}$ $\pi$ & -18.09  & -17.98 & -18.00 & -25.65 \\
 1s$_{1/2}$ $\nu$ & -40.55  & -40.62 & -40.61 & -42.21 \\
 1p$_{3/2}$ $\nu$ & -15.05  & -14.76 & -14.81 & -25.26 \\
 1p$_{1/2}$ $\nu$ & -15.05  & -15.77 & -15.66 & -16.91 \\
\hline
\end{tabular}
\end{center}
\end{table}
\begin{table}
\caption{Same as table \ref{tablec12} for $^{16}$O.}
\label{tableo16}
\begin{center}
\begin{tabular}{crrrr}
\hline
 $^{16}$O   & B1      & JSB    &   A    & T6    \\ 
\hline
 1s$_{1/2}$ & -49.40  & -49.55 & -49.55 & -49.39 \\
 1p$_{3/2}$ & -21.71  & -21.79 & -21.79 & -21.71 \\
 1p$_{1/2}$ & -21.71  & -21.79 & -21.79 & -21.71 \\
\hline
\end{tabular}
\end{center}
\end{table}
\begin{table}
\caption{Same as table \ref{tablec12} for $^{40}$Ca.}
\label{tableca40}
\begin{center}
\begin{tabular}{crrrr}
\hline
 $^{40}$Ca   & B1      & JSB    &   A    &  T6\\ 
\hline
 1s$_{1/2}$  & -71.34  & -71.57 & -71.57 & -71.34  \\
 1p$_{3/2}$  & -45.24  & -45.39 & -45.39 & -45.24 \\
 1p$_{1/2}$  & -45.24  & -45.39 & -45.39 & -45.24 \\ 
 1d$_{5/2}$  & -21.19  & -22.04 & -22.04 & -21.95  \\
 1d$_{3/2}$  & -21.19  & -22.04 & -22.04 & -21.95  \\
 2s$_{1/2}$  & -20.17  & -20.26 & -20.26 & -20.17  \\
\hline
\end{tabular}
\end{center}
\end{table}
\begin{table}
\caption{Same as table \ref{tablec14} for $^{48}$Ca.}
\label{tableca48}
\begin{center}
\begin{tabular}{crrrr}
\hline
 $^{48}$Ca        & B1      & JSB    &   A    & T6 \\ 
\hline
 1s$_{1/2}$ $\pi$ & -73.87  & -74.08 & -74.07 & -74.43 \\
 1p$_{3/2}$ $\pi$ & -50.81  & -50.56 & -50.53 & -51.76 \\
 1p$_{1/2}$ $\pi$ & -50.81  & -51.03 & -51.06 & -51.04 \\  
 1d$_{5/2}$ $\pi$ & -28.95  & -28.72 & -28.70 & -34.07 \\
 1d$_{3/2}$ $\pi$ & -28.95  & -29.50 & -29.50 & -22.45 \\
 2s$_{1/2}$ $\pi$ & -26.96  & -27.54 & -27.48 & -26.96 \\
 1s$_{1/2}$ $\nu$ & -69.57  & -69.78 & -69.77 & -70.06 \\
 1p$_{3/2}$ $\nu$ & -45.44  & -45.49 & -45.47 & -44.96 \\
 1p$_{1/2}$ $\nu$ & -45.98  & -46.27 & -46.29 & -48.87 \\ 
 1d$_{5/2}$ $\nu$ & -23.64  & -23.57 & -23.55 & -25.87 \\
 1d$_{3/2}$ $\nu$ & -24.47  & -24.78 & -24.79 & -21.74 \\
 2s$_{1/2}$ $\nu$ & -22.94  & -23.03 & -23.29 & -22.88 \\
 2f$_{7/2}$ $\nu$ & -4.19   &  -4.06 &  -4.05 &  -7.57 \\
\hline
\end{tabular}
\end{center}
\end{table}
\clearpage
{\bf Figure Captions}
\begin{figure} [h]
\caption{   
Tensor isospin terms added to the B1 interaction. 
In the panel (a) the curve labeled
with A represents the term of the 
Argonne $v_{14}$ potential \protect\cite{wir84}, that labeled as JSB
the term of the J\"ulich--Stony Brook interaction of
ref. \protect\cite{spe80}. The panel (b) shows the interaction
adjusted to reproduce the $^{12}$C experimental binding energy. 
 }
\label{fig1}
\end{figure}
\begin{figure} [h]
\caption{   Proton density distributions 
obtained with the B1 interaction (full lines) with the A interaction 
(dashed lines), with the JSB interaction (dotted--dashed lines
and with the T6 interaction (doubly dotted--dashed lines).
On the scale of the figure the A and JSB results are overlapped.
 }
\label{fig2}
\end{figure}
%
%\end{document}
% ---------------------------------------
%          include ps figures
% ---------------------------------------
\clearpage
\begin{figure}[t]
\vspace{-7cm}
\begin{center}                                                                
\leavevmode
\epsfysize = 750pt
\makebox[0cm]{\epsfbox{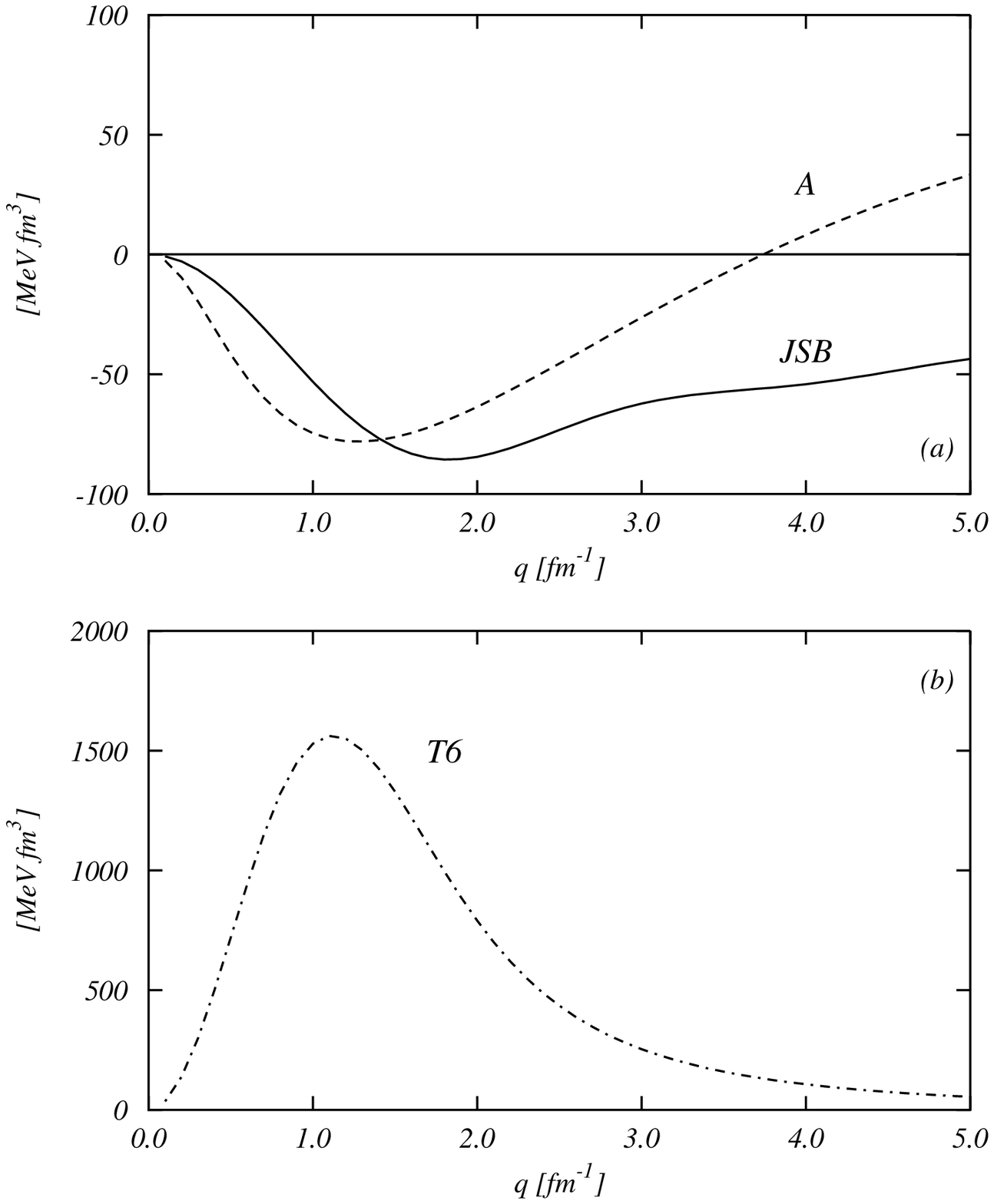}}
\end{center}
%\vspace{-5.5cm}
\end{figure}
\begin{figure}[t]
\vspace{-2cm}
\begin{center}                                                                
\leavevmode
\epsfysize = 600pt
\makebox[0cm]{\epsfbox{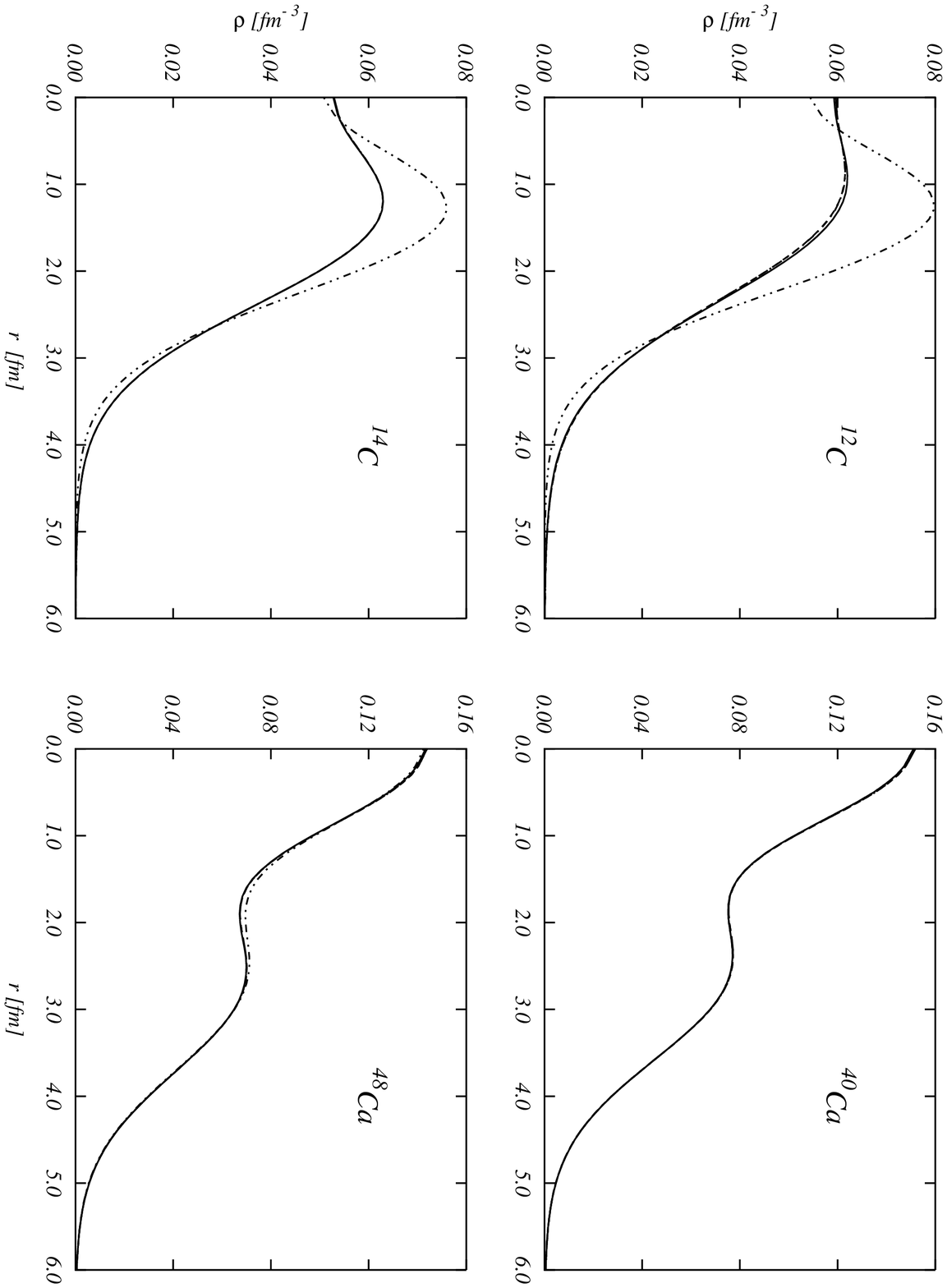}}
\end{center}
%\vspace{-5.5cm}
\end{figure}

\end{document}